\documentclass{article}

\usepackage{arxiv}

\usepackage[utf8]{inputenc} 
\usepackage[T1]{fontenc}    
\usepackage{hyperref}       
\usepackage{url}            
\usepackage{booktabs}       
\usepackage{amsfonts}       
\usepackage{nicefrac}       
\usepackage{microtype}      
\usepackage{lipsum}
\usepackage{graphicx}
\graphicspath{ {./images/} }

\title{Exploring AI Writers: Technology, Impact, and Future Prospects}

\author{
  Zhiqian Huang \\
  fanqienovel.com\\
  Baoshan Qu, Shanghai \\
  \texttt{15038073597@163.com} \\
}

\begin{document}
\maketitle
\begin{abstract}
  This study explores the practical capabilities of AI writers, focusing on their 
  applications across various creative domains. It delves into the potential impact of AI-generated content on traditional media industries and academic writing processes. The research examines how AI tools are reshaping news production workflows, particularly in fields such as finance, sports, and natural disasters. Additionally, it addresses ethical concerns, including authorship and copyright issues arising from AI-driven creative outputs. The findings reveal mixed perceptions among media students regarding the integration of AI into their profession, reflecting both optimism about efficiency gains and apprehensions over increased job market competition.
\end{abstract}


\section{Introduction}
Artificial Intelligence (AI) writers have emerged as a significant force in the realm of content creation. These advanced tools leverage natural language processing techniques to generate coherent and logical texts, applicable across various domains such as journalism, advertising, and educational materials. This document delves into the capabilities, applications, and implications of AI writers, examining their technological underpinnings, market influence, strengths, limitations, future trajectories, and ethical considerations.

In the rapidly evolving landscape of artificial intelligence technologies today, AI models are increasingly being applied across various domains, with literary creation being no exception. Recently, We had the privilege of reading five insightful articles that not only explored the applications of AI in academic writing and interactive artistic creation but also delved into the profound ethical and copyright issues raised by AI-generated content, providing us with invaluable pathways for reflection.

\subsection{the theory of cognitive psychology} 
Based on cognitive psychology theories, this study focuses on users’ cognitive needs by exploring interactions between users and AI writing assistants within AI writing editors. It aims to provide theoretical guidance for the future development of AI writing editor interfaces. The methodology involves an in-depth analysis of behavioral differences between determinate and indeterminate creators, as well as cognitive characteristics such as visual aesthetics perception, attention span, learning and memory capabilities, and cognitive thinking patterns. Key issues identified include information overload leading to cognitive strain, conflicts between content and short-term memory storage limits, weak AI entry point awareness, and difficulties in user recall.\cite{1}

This article primarily investigates the influence of AI-assisted writing feedback on second-language learners’ writing performance, self-efficacy in writing, and writing-related self-regulation. It emphasizes the educational effects of AI-generated writing feedback. In the domain of foreign language education, advancements in artificial intelligence are driving transformations in the foreign language classroom. Against this backdrop, foreign language education is exhibiting trends toward personalized, intelligent, and pervasive integration of AI technologies. Within foreign language writing instruction, AI chatbots have introduced new research topics for second-language writing feedback studies. While computer-assisted feedback in second-language writing has garnered significant attention, research on AI-assisted second-language writing feedback remains relatively limited. Thus, this study aims to explore the impacts of AI-assisted writing feedback on students’ writing outcomes, self-efficacy in writing, and writing-related self-regulation.

This article approaches its inquiry from both industry and educational perspectives, focusing on the implications of AI writing for the structure of the media industry and talent development. It highlights how AI writing has transformed actual workflows within the media sector while also addressing how education should cultivate professionals adapted to the AI era in journalism. The rise of machine learning-based news writing has seen extensive application across domains such as finance, sports, and natural disasters, reshaping the internal production structures of the media industry from both temporal and spatial dimensions. In response to these developments, media students generally hold a cautious yet hopeful outlook on the impact of AI on the journalism profession, perceiving increased employment pressure for journalists while also anticipating significant improvements in news content production efficiency. Amid the transformative winds of AI in the media sector, government agencies, media organizations, educational institutions, and individual students are urged to align with these shifts collectively and foster versatile talent suited to the new environments of the evolving journalism industry.
\subsection{foreign language education}
In the field of foreign language education, the development of artificial intelligence (AI) is leading the transformation of foreign language classrooms. Against this backdrop, foreign language education is presenting AI-driven trends that are individualized, intelligent, and widely popularized. In the direction of foreign language writing instruction, AI chatbots have introduced new research topics for second language writing feedback studies. While computer-assisted second language writing feedback has received extensive attention, research on AI-assisted second language writing feedback remains relatively limited. Therefore, this study aims to explore the impact of AI-assisted writing feedback on students' writing performance, their sense of self-efficacy in writing, and their self-regulation in writing. This study attempts to answer the following three research questions:\cite{2}  
1. What is the impact of AI-assisted writing feedback on second language learners' writing performance?  
2. What is the impact of AI-assisted writing feedback on second language learners' sense of self-efficacy in writing?  
3. What is the impact of AI-assisted writing feedback on second language learners' self-regulation in writing?  

This article identifies key issues affecting users' cognitive processes in interface design by analyzing differences in behavior between deterministic and non-deterministic creators, as well as users' visual aesthetics, attention, learning, memory, and cognitive thinking characteristics.

\subsection{Application in other fileds}

Artificial intelligence (AI) news writing has been widely applied in news topics such as finance, sports, and natural disasters, reshaping the internal news production structure of the media industry from both temporal and spatial perspectives. In response to this trend, many media students hold a cautious optimism about the impact of AI on the media industry, believing it could increase employment pressure on media professionals while also holding high expectations for its potential to enhance news content production efficiency. Amidst the wave of AI's impact on the media industry, government departments, media companies, universities, and students themselves must keep pace with the times and collectively work towards fostering composite talents adapted to the new environment of the media industry.\cite{3} 

This article approaches the topic from both industry and educational perspectives, focusing on the implications of AI writing for the structure of the media industry and talent development. It emphasizes how AI writing is transforming the actual workflows within the media industry while also addressing how education can cultivate media professionals suited to the AI era.

\subsection{AI Transforming Media Logic and Production Efficiency}
Artificial Intelligence (AI) is driving significant transformations across the globe. Within the media sector, AI is set to redefine media logic. Machine writing and speech recognition, among other AI technologies, have substantially improved media content production efficiency, allowing media professionals to allocate more energy toward contemplating and creating deeper, more meaningful content. In the AI era, unlike the previous creator-oriented or user-oriented approaches, the new media logic will be technology-driven.\cite{4}  

This article emphasizes how AI technologies are reshaping media logic and boosting productivity. It highlights the fundamental shift AI is bringing about in the production models of the media industry while also underscoring how these technologies enable professionals to focus on producing more substantial and impactful content.

\subsection{AI Writing vs. Human Writing: Exploring the Potential of LLM tools}
One of the defining features of the current era is the acceleration of technological advancements, which has simultaneously led to a decline in the primacy of human intelligence. As writing, a uniquely human cultural behavior, undergoes significant transformations, AI writing is emerging as a new transformative force. However, there is an inherent difference between AI-driven writing and human writing; the former appears to freely manipulate large datasets but is essentially constrained by programming rules, while the latter embodies real experiences and emotions, reflecting the true essence of human embodiment in writing—a crucial means of realizing human values and meanings.\cite{5}

This article adopts a critical perspective on AI writing, emphasizing the unique value and significance of human writing. It points out that AI writing is merely an imitation within programmed constraints, whereas human writing encapsulates genuine experiences and emotions, serving as an effective vehicle for realizing human values and meanings. This is indeed one of the most insightful interpretations.

However, the discussions above primarily focus on the perspectives of professionals and have limited exploration into the application scope of AI in specific functions, particularly in the niche domain of novel creation. Given this, as a novelist myself, I deeply feel the necessity to delve deeper from a creator's viewpoint, analyzing the characteristics and potential of three prominent models—Wenxin Yiyan, DaoBei Fang, and DeepSeek—in novel writing.

Novels, as an important literary form, serve multiple functions: narrating stories, developing characters, conveying emotions, while also embodying authors' unique aesthetic concepts and writing styles. The emergence of AI models undoubtedly brings new possibilities and challenges to novel creation.

This paper aims to explore the distinctive features and advantages of these three AI models in aspects such as plot development, character portrayal, and language style through comparative analysis of their application instances in novel writing. I hope that this research will provide practical references for peers regarding AI applications in novel creation and contribute to the exploration of integrating AI with literary creativity.

\section{Function and Application of AI Writers}
AI writers are capable of producing a wide array of content, from simple articles to complex novels. They can generate texts based on user input regarding themes, formats, and styles. 
Here we have briefly written a reference for the inspiration of a novel based on us usual writing experience, and the following is a prompt:

\subsection*{Content Generation}

 Title : *The Elegant Dance of the Sword in the World of Rivers and Lakes*  
 Main Plot Thread : Luo Yi, a man who appears ordinary but is actually extraordinary, calm, and introverted. He has two sisters, Bao Xue and Bao Ya. The former is gentle and refined, while the latter is cold and elegant, both bound to him by fate. Luo Yi intends to live an uneventful life in this ancient wuxia world, but becomes entangled in events like rivers and lakes disputes, faction strife, and resource struggles due to various reasons. He faces numerous hardships, such as being jilted, poisoned, and suffering from amnesia, yet grows continuously, eventually reaching a state of solitary failure and joining his sisters in reshaping the world's principles.  

 Tags :  
-  Character Setup : Proud and lonely invincible man and Two life-and-death sisters vs Various opposing factions  
-  Plot Points : Invincible flow, wuxia, multiple female leads, eating pork and tiger meat  
-  Era : Ancient setting  
-  Genre : Wuxia  
-  Target Audience : Male frequency  

 Background Setting :  
-  Protagonist's Golden Fingers : Luo Yi possesses a unique inner cultivation method that doubles his training efficiency and automatically repels heart demons and temptations.  
-  Ranking System :  (society) are divided into levels such as newly joined, small reputation, named in one region, worldly skilled, and transcendental saint.  

 Character Design :  
-  Main Strengths :  
  - Luo Yi (male lead, wandering swordsman, calm and introverted but actually powerful)  
  - Bao Xue (first female lead, eldest of the Bao family, gentle and refined, fate-bound with male lead)  
  - Bao Ya (second female lead, younger sister of Bao family, cold and elegant, changing from cold to hot for male lead)  

 Early Phase   
-  Storyline : Luo Yi reveals extraordinary abilities when provoked in his small town.  
-  Climax : Luo Yi easily defeats the local villain, whose so-called top fighter is also suppressed by him.  

 Other Characters :  
-  Lin Bo  (supporting character, kind-hearted townsfolk)  
-  Zhan Tiejun  (supporting character, town blacksmith, generous and just)  
-  Sunu Nü  (supporting character, town doctor, kind and wise)  
-  Chen Xiucai  (supporting character, town scholar, eccentric but upright)  

 Mid Phase   
-  Main Storyline : Luo Yi gets involved in resource struggles between the Bao family and other factions, with Bao sisters starting to follow him.  
-  Climax : Luo Yi discerns various conspiracies during the struggle for treasures, defeating multiple enemies at once.  

 Other Characters :  
-  Qin Ronglao  (antagonist, leader of a certain faction, pretending kindness but plotting against Luo Yi)  
-  Jue Kuang  (antagonist, ruthless martial artist, hired to kill Luo Yi)  
-  Li Tangzhu  (supporting character, loyal and fierce Bao subordinate)  
-  Zhou Xiansheng  (supporting character, Bao strategist, wise and scheming)  

 Late Phase   
-  Main Storyline : A great calamity approaches, Luo Yi and the Bao sisters jointly seek a resolution.  
-  Climax : Luo Yi refines his cultivation method to match treasures and joins the Bao sisters in reshaping world principles.  

 Other Characters :  
-  Ying Mo  (antagonist, demon or ghost plotting chaos)  
-  Xi Menxiaoxue  (antagonist, evil sorcerer opposing Luo Yi)  
-  Hua Shenyi  (supporting character, divine doctor aiding Luo Yi and others)  
-  Song Laozhang  (supporting character, reclusive sect leader guiding Luo Yi)  

 Main Emotive Elements, Turning Points, and Reader Hooks :  
- ( Surprising ) Luo Yi's tiger meat-eating reversal.  
- ( Tragic ) Tragic events involving the Bao sisters.  
- ( Anticipation ) How Luo Yi navigates various conspiracies.  
- ( Surprise ) Artifact ownership and strange encounters.



\section{Comparison and analysis}
Let's first analyze the beginning of the three model creations:

Wen Xin: In the afternoon of the town, the sun shines like a filament, penetrating the sparse clouds and shining on the bluestone road. There are few pedestrians on the street, and occasionally there are a few children's laughter, which adds a bit of vitality to this quiet picture. Luo Yi, a young man in old clothes and a bamboo hat, walks slowly on the street. His pace is neither fast nor slow, as if every step is in a certain rhythm, blending with the tranquility of the town.

Dou Bao: In the early morning, the town is gently wrapped in a layer of light mist, and the stalls in the market seem to wake up slowly from sleep and open one by one. Luo Yi walks towards the market at a slow pace as usual. His steps are leisurely, but with a unique leisurely feeling. deepseek: In the tavern in the town, the sun shines on the mottled wooden table through the half-open window, and the air is filled with a faint aroma of wine and fireworks. Luo Yi lazily leaned on a chair in the corner, holding a glass of sake in his hand, his eyes blurred, as if he was indifferent to everything around him. His hair was a little messy, his clothes were a little old, and he looked like an ordinary lazy teenager, out of tune with the other busy figures in the tavern.

Wenxin's writing style is more delicate, and he naturally used the metaphor "sunlight sprinkles like filaments" at the beginning. This style is suitable for depicting delicate emotions and tranquil scenes.

Doubao's writing style is more vivid and dynamic, like "the town is gently wrapped in a layer of light mist", which is suitable for depicting dynamic scenes and characters' activities.

Deepseek's writing style is more rough and realistic, focusing on the true presentation of details, and describing the interior environment of the tavern in detail, including sunshine, wooden tables, wine aroma and fireworks. Luo Yi's image is more three-dimensional here. His lazy, indifferent attitude, messy hair and old clothes all form a sharp contrast with the tavern environment he is in, naturally showing a unique and somewhat decadent image of a teenager.

Next, let's take a look at the differences in the conflict parts of the three model creation stories:

Wen Xin: He was thinking, and the two men suddenly stopped. One of them pointed at Luo Yi and shouted: "Hey, kid! Did you see your uncle run away? Do you want to meddle in other people's business?" Luo Yi was stunned, then shook his head, saying that he was just passing by. But the man was relentless, stepped forward and pushed Luo Yi: "Humph, I think you are guilty! Tell me, did you steal our purse?" Luo Yi frowned. He didn't want to cause trouble, but the two were too rude. He was about to explain, but the man couldn't bear it anymore and punched him. Luo Yi's body suddenly shook slightly, as if a gust of wind blew. The next moment, the man's fist had landed in the air, but his hand was firmly grasped by Luo Yi, unable to move. The man's eyes widened, his face full of disbelief: "You...how could you..." Luo Yi smiled faintly and said, "I just don't want to cause trouble, but it doesn't mean I'm afraid of trouble. You'd better be honest and don't act wild here."

Bean bun: Just as Luo Yi was enjoying the delicious snacks, a burst of noise suddenly broke the peace. Wang Mazi came over with a group of men, and their footsteps sounded particularly harsh on the stone road. Wang Mazi walked to Miss Su's medicine stall, and without saying a word, he kicked over the medicine stall, and the medicinal materials scattered all over the ground. "You quack, the medicine you prescribed for me last time was useless, you must give me an explanation today!" Wang Mazi shouted at the top of his voice, with a hint of arrogance and arrogance on his face. Miss Su was angry and anxious, her eyes instantly turned red, and her voice trembled a little: "The medicine I prescribed is all for the symptoms. It was you who didn't follow the doctor's advice and drank and ate meat every day. How can you blame me for the ineffectiveness of the medicine?" Seeing this, Luo Yi frowned tightly into a "chuan" character, walked forward a few steps, and stared at Wang Mazi coldly: "Wang Mazi, bullying a girl, do you still have shame?" When Wang Mazi saw Luo Yi, a trace of ferocity appeared on his face, and a mocking smile appeared at the corner of his mouth: "Luo Yi, don't think that I am afraid of you just because you pretend to be honest. Don't bother yourself with this matter today! Get out of here if you know what's good for you!" As he spoke, he rolled up his sleeves and made a gesture to fight. deepseek: "Boy, I heard that you are quite arrogant recently? Dare you not take me seriously?" The bully's voice was rough and threatening, and the atmosphere in the tavern instantly became tense. The guests lowered their heads one after another, not daring to look at the bully, for fear of getting into trouble. Luo Yi raised his head, glanced at the bully lazily, and said lightly: "You are blocking my light." As soon as this sentence came out, the tavern was suddenly silent. The bully's face turned livid instantly. He didn't expect that this ordinary-looking boy would dare to contradict him like this. He slammed the table and roared: "Teach him a lesson!" Several men surrounded him with a grin, and their fists clenched and creaked. The guests in the tavern retreated one after another, fearing that they would be affected. Lin Bo and Zhao the blacksmith also stood up, ready to help, but were stopped by Luo Yi's eyes. At the moment when the bully's men were about to take action, Luo Yi's eyes flashed with a cold light. He still sat on the chair, just gently raised his right hand, and flicked his fingertips slightly. An invisible energy burst out instantly, and the bully's men were struck by lightning, flying backwards, knocking over tables and chairs, and falling to the ground in embarrassment. In this conflict plot, Wen Xin quickly unfolded a small episode through the conflict between Luo Yi and the two men. The rudeness of the man and Luo Yi's response constitute the main conflict point of the plot. The language is concise and clear, and the dialogue is natural and smooth, successfully creating a tense and explosive atmosphere.

Doubao's plot is richer, involving multiple characters such as Luo Yi, Miss Su and Wang Mazi, as well as the complex relationship between them. Wang Mazi's arrogance and Miss Su's helplessness constitute the main conflict point of the plot, and Luo Yi's intervention makes the plot more tortuous.

Deepseek's plot is more dramatic, and a thrilling episode unfolds through the bully's threat and Luo Yi's counterattack. The bully's arrogance and Luo Yi's calmness constitute the main conflict point of the plot. Deepseek's language is more concise and full of tension. Through the bully's roar, Luo Yi's light words and the description of his men being struck by lightning, readers can feel a strong visual and emotional impact.

\subsection{analysis}

The analysis below compares three different writing styles (Wenxin, Doubao, and Deepseek) in terms of their opening sections and conflict parts within a narrative. Each model exhibits unique characteristics that distinguish its approach to storytelling.
---

\subsubsection{Opening Sections Analysis}

-  Wenxin : The opening is delicate and subtle, using metaphors like "sunlight as soft as silk." This style suits the depiction of refined emotions and serene scenes.  
-  Doubao : The writing is more dynamic and vivid, exemplified by phrases like "the town covered in a thin layer of fog." This approach is suitable for depicting lively scenarios and character actions.  
-  Deepseek : The style is rougher and realistic, emphasizing details such as sunlight through windows and the atmosphere of a wine shop. It focuses on presenting an authentic environment and distinct characters.

---

\subsubsection{Conflict Parts Analysis}

-  Wenxin : The conflict involves a direct encounter between Luo Yi and two men, with straightforward language and natural dialogue, creating a tense yet fluid atmosphere.  
-  Doubao : The narrative includes multiple characters like Luo Yi, Su Guniang, and Wang Mazi, showcasing complex relationships that add twists to the plot.  
-  Deepseek : The conflict is more dramatic, featuring Eviluo's intimidation and Luo Yi's calm response, culminating in a surprising turn of events. The language is concise and impactful.

---

\subsection{summary}

Each model Wenxin, Doubao, and Deepseek demonstrates distinct strengths in storytelling. Wenxin excels at subtle emotion and serenity, Doubao at dynamic action, and Deepseek at dramatic intensity. Their differences highlight various approaches to narrative construction.

---

\section{Conclusion}

\bibliographystyle{unsrt}  


\end{document}